\documentclass[12pt]{article}
\usepackage{amsmath}  
\input amssym

\begin{document}
\def\prg#1{\medskip{\bf #1}}     \def\ra{\rightarrow}
\def\lra{\leftrightarrow}        \def\Ra{\Rightarrow}
\def\nin{\noindent}              \def\pd{\partial}
\def\dis{\displaystyle}          \def\inn{\,\rfloor\,}
\def\grl{{GR$_\Lambda$}}         \def\vsm{\vspace{-9pt}}
\def\Lra{{\Leftrightarrow}}      \def\btz{{\rm BTZ}}
\def\cs{{\scriptstyle\rm CS}}    \def\ads3{{\rm AdS$_3$}}
\def\mb{{\scriptstyle\rm MB}}    \def\tmg{{\scriptstyle\rm TMG}}
\def\ric{{(Ric)}}                \def\tmgl{\hbox{TMG$_\Lambda$}}
\def\Lie{{\cal L}\hspace{-.7em}\raise.25ex\hbox{--}\hspace{.2em}}

\def\G{\Gamma}        \def\S{\Sigma}        \def\L{{\mit\Lambda}}
\def\D{\Delta}        \def\Om{\Omega}
\def\a{\alpha}        \def\b{\beta}         \def\g{\gamma}
\def\d{\delta}        \def\m{\mu}           \def\n{\nu}
\def\th{\theta}       \def\k{\kappa}        \def\l{\lambda}
\def\vphi{\varphi}    \def\ve{\varepsilon}  \def\p{\pi}
\def\r{\rho}          \def\om{\omega}       \def\s{\sigma}
\def\t{\tau}          \def\eps{\epsilon}    \def\nab{\nabla}

\def\bA{{\bar A}}     \def\bF{{\bar F}}     \def\bG{{\bar G}}
\def\bt{{\bar\tau}}   \def\bg{{\bar g}}     \def\tG{{\tilde G}}
\def\cL{{\cal L}}     \def\cM{{\cal M }}    \def\cE{{\cal E}}
\def\cH{{\cal H}}     \def\bcH{{\bar\cH}}   \def\hcH{\hat{\cH}}
\def\cK{{\cal K}}     \def\hcK{\hat{\cK}}
\def\cO{{\cal O}}     \def\hcO{\hat{\cal O}}
\def\cB{{\cal B}}     \def\heps{\hat\epsilon}

\def\br{{\bar\r}}     \def\bb{{\bar b}}     \def\bt{{\bar\t}}
\def\bpi{{\bar\pi}}   \def\bom{{\bar\om}}   \def\bphi{{\bar\phi}}
\def\tom{{\tilde\omega}} \def\tR{{\tilde R}}
\def\nn{\nonumber}
\def\be{\begin{equation}}             \def\ee{\end{equation}}
\def\ba#1{\begin{array}{#1}}          \def\ea{\end{array}}
\def\bea{\begin{eqnarray} }           \def\eea{\end{eqnarray} }
\def\beann{\begin{eqnarray*} }        \def\eeann{\end{eqnarray*} }
\def\beal{\begin{eqalign}}            \def\eeal{\end{eqalign}}
\def\lab#1{\label{eq:#1}}             \def\eq#1{(\ref{eq:#1})}
\def\bsubeq{\begin{subequations}}     \def\esubeq{\end{subequations}}
\def\bitem{\begin{itemize}}           \def\eitem{\end{itemize}}
\renewcommand{\theequation}{\thesection.\arabic{equation}}

\title{Conserved charges in 3D gravity} 

\author{M. Blagojevi\'c and B. Cvetkovi\'c\footnote{
        Email addresses: {\tt mb@ipb.ac.rs,
                                cbranislav@ipb.ac.rs}} \\
University of Belgrade, Institute of Physics,\\ P. O. Box 57, 11001
Belgrade, Serbia}
\date{}
\maketitle
\begin{abstract}
The covariant canonical expression for the conserved charges, proposed
by Nester, is tested on several solutions in three-dimensional gravity with or without
torsion and topologically massive gravity. In each of these cases, the
calculated values of energy-momentum and angular momentum are found to
satisfy the first law of black hole thermodynamics.
\end{abstract}

\section{Introduction}

Topologically massive gravity with a cosmological constant (\tmgl) is
an extension of the usual, Einstein's three-dimensional (3D) gravity,
obtained by adding the gravitational Chern-Simons term to the action
\cite{1}. This step leads to an essential modification of Einstein's
(topological) theory, giving rise to a propagating degree of freedom,
the massive graviton. However, for generic values of the coupling
constants, the physical interpretation of \tmgl\ in the anti-de Sitter (AdS) sector
suffers from serious difficulties, related to the instability of the
\ads3\ vacuum \cite{1,2,3}. To avoid the problem, Li et al. \cite{3}
introduced the so-called chiral version of \tmgl\ and argued that it
might be free from these difficulties, but there are also opposite
opinions \cite{4}. Another option would be to choose a new vacuum, such
as the spacelike stretched \ads3, which could be a stable ground state
of the theory \cite{5}.

It is clear that all these issues are closely related to the
concept of conserved charges, which is essential for a clear
understanding of the stability problem. The conserved charges were
calculated for a number of different solutions of \tmgl\
\cite{6,7,8,9}, but, as noted in \cite{10}, the methods used in
these papers are not best suited for studying general structure of
the boundary dynamics. An attempt to improve the situation was
made by Mi\v skovi\'c and Olea \cite{10}, who proposed a
background-independent Noether construction of the conserved
charges in \tmgl. They found correct conserved charges for the
Ba\~nados-Teitelboim-Zanelli (BTZ)
black hole and the logarithmic solution; however, when applied to
the spacelike stretched black hole, their method yields the value
of energy with an extra factor $1/2$, as compared to \cite{7,9},
the result that is not supported by the first law of black hole
thermodynamics \cite{7,5}. In the present paper, we focus our
attention on another idea---the {\it covariant canonical
formalism\/} developed by Nester \cite{11} and collaborators
\cite{12,13,14,15}, which has had excellent results for a large
class of the gravitational solutions in 4D. We intend to show that
it leads to a rather general expression for the conserved charges
in 3D gravity, valid not only in \tmgl but also in 3D gravity
with or without torsion.

The paper is organized as follows. In section 2, we give a brief
account of the general first order Lagrangian formalism in 3D gravity,
with a focus on (a) 3D gravity with or without torsion and (b) \tmgl.
In section 3, we introduce Nester's covariant canonical formalism and
define general covariant expressions for the conserved charges. In
section 4, we use these results to evaluate the conserved charges for
several solutions in 3D gravity, including not only the BTZ black hole
and the logarithmic solution, but also the spacelike stretched black
hole. The conserved charges found here satisfy the first law of
thermodynamics, and moreover, they are in agreement with the results
obtained earlier in \cite{7,8,9,16,17}. Section 5 is devoted to
concluding remarks, and appendices contain some technical details.

Our conventions are as follows: the Latin indices $(i,j,k,...)$ refer
to the local Lorentz frame, the Greek indices $(\m,\n,\l,...)$ refer to
the coordinate frame, and both run over 0,1,2; the metric components in
the local Lorentz frame are $\eta_{ij}=(+,-,-)$; totally antisymmetric
tensor $\ve^{ijk}$ is normalized to $\ve^{012}=1$. Our notation follows
the Poincar\'e gauge theory (PGT) framework in 3D \cite{8,9}:
fundamental dynamical variables are the triad field $b^i$ and the
Lorentz connection $\om^i$ (1-forms), $T^i=db^i+\ve^{ijk}\om_j b_k$ and
$R^i=d\om^i+\frac{1}{2}\ve^i{}_{jk}\om^j\om^k$ are the corresponding
field strengths, the torsion and the curvature (2-forms), the wedge
product signs ($\wedge$) are omitted for simplicity and the relation to
the standard 4D notation is given by $\om^{ij}=-\ve^{ij}{_k}\om^k,
R^{ij}=-\ve^{ij}{_k}R^k$.

\section{First order Lagrangians in 3D gravity}
\setcounter{equation}{0}

Using Nester's ideas \cite{11}, we show in this section that the
following gravitational models in three dimensions:
\bitem
\item[(a)] 3D gravity with torsion \cite{18,16}, including also Einstein's
           3D gravity, and\vsm
\item[(b)] topologically massive gravity \cite{1},
\eitem
can be described by a generic {\it first order\/} Lagrangian (density),
with
\be
\cL=\t_i T^i+\r_i R^i-V(b^i,\om^i,\t_i,\r_i)\, .           \lab{2.1}
\ee
Here, not only $b^i$ and $\om^i$, but also $\t^i$ and $\r^i$ are {\it
independent\/} dynamical variables (1-forms), and $V$ is a conveniently
chosen term for each of the above two cases, as described below.

(a) Three-dimensional gravity with torsion is a topological theory in
{\it Riemann-Cartan spacetime,\/} defined by the Mielke-Baekler
Lagrangian \cite{18,16}:
\bsubeq\lab{2.2}
\be
\cL_\mb=2ab^i R_i-\frac{\L}{3}\,\ve_{ijk}b^ib^jb^k\,
        +\a_3 L_\cs(\om)+\a_4 b^i T_i\, ,                  \lab{2.2a}
\ee
where $a=1/16\pi G$, and $L_\cs(\om)=\om^i d\om_i
+\frac{1}{3}\ve_{ijk}\om^i\om^j\om^k$ is the Chern-Simons Lagrangian
for the Lorentz connection. The Lagrangian can be rewritten in the form
\be
\cL_\mb=\t_i T^i+\r_i R^i -\frac{\L}{3}\,\ve_{ijk}b^ib^jb^k
        -\frac{\a_3}6\ve_{ijk}\om^i\om^j\om^k\, ,          \lab{2.2b}
\ee
where $\t_i$ and $\r_i$ are defined as the covariant field momenta,
conjugate to the field strengths $T^i$ and $R^i$:
\be
\t_i:= \frac{\pd L_\mb}{\pd T^i}=\a_4 b_i\, ,\qquad
\r_i:= \frac{\pd L_\mb}{\pd R^i}= 2ab_i+\a_3\om_i\,.       \lab{2.2c}
\ee
\esubeq

Now, instead of using the original Lagrangian \eq{2.2}, it is more
convenient to reformulate it so that $\t_i$ and $\r_i$ become {\it
independent\/} dynamical variables. This goal is achieved by the first
order Lagrangian \eq{2.1}, where the constraints \eq{2.2c} are enforced
via Lagrange multipliers in $V$ \cite{12,14}.

Note that for $\a_3=\a_4=0$, the Mielke-Baekler Lagrangian reduces to
Einstein's 3D gravity (with a cosmological constant) in Riemannian
spacetime. Thus, the first order Lagrangian \eq{2.1} is suitable also
for Einstein's 3D gravity.

(b) A simple description of \tmgl\ is given by the Lagrangian \cite{8}
\bsubeq
\be
\cL_\tmg=2ab^i R_i-\frac{\L}{3}\,\ve_{ijk}b^ib^jb^k\,
         +a\m^{-1}L_\cs(\om)+\l^i T_i\, ,                  \lab{2.3a}
\ee
where $\l^i$ (1-form) is a Lagrange multiplier that ensures the
vanishing of torsion. This theory can  also be transformed into the form
\eq{2.1}, with a choice of $V$ that ensures the correct on-shell values
of $\t_i,\r_i$:
\be
\t_i\approx\frac{\pd L_\tmg}{\pd T^i}=\l_i\, ,\qquad
\r_i\approx\frac{\pd L_\tmg}{\pd R^i}= 2ab_i+a\m^{-1}\om_i\,.\lab{2.3b}
\ee
\esubeq

The explicit form of $V$ in the generic Lagrangian \eq{2.1} is not
important for our discussion of the conserved charges; all we need to
remember is that $V$ {\it does not contain derivatives of dynamical
variables,\/} $V=V(b^i,\om^i,\t_i,\r_i)$.

\section{Covariant canonical formalism for conserved charges}
\setcounter{equation}{0}

Starting with the first order Lagrangian \eq{2.1} and using the
field-theoretic analogue of the classical mechanics relation
$Ldt=(p\dot q-H)dt$, one can introduce a timelike vector field $\xi$
and define the Hamiltonian 2-form $\cH(\xi)$ on the spatial
hypersurface $\S$ of spacetime \cite{11}. In analogy with old Dirac's
ideas \cite{19}, these considerations can be generalized by allowing
$\xi$ to be either timelike {\it or\/} spacelike, whereupon $\cH(\xi)$
becomes the {\it generalized Hamiltonian\/} density, associated with the
dynamical evolution along $\xi$.

The Hamiltonian density $\cH(\xi)$ contains a boundary term $dB$, but
the requirement that $\cH(\xi)$ generates the correct equations of
motion allows a freedom in the choice of $B$. As shown by Regge and
Teitelboim \cite{20}, the proper form of $B$ is determined by the
requirement that the functional derivatives of $H(\xi)=\int_\S\cH(\xi)$
are well defined or, equivalently, that the boundary term in $\d
H(\xi)$ vanishes. The verification of this criterion is closely related
to the specific form of the asymptotic conditions.

If $\xi$ is asymptotically a Killing vector, the related conserved
charge is naturally identified as the on-shell value of $H(\xi)$. Since
the generalized Hamiltonian has the form
$$
H(\xi)=\int_\S \xi^\m\cH_\m+\int_{\pd\S}B(\xi)\, ,
$$
where $\cH_\m$ is found to be proportional to the field equations,
the value of $H(\xi)$ on a solution of the field equations reduces
to the boundary integral $\int_{\pd\S} B(\xi)$. Hence, the corresponding
conserved charge is defined as the value of $\int_{\pd\S} B(\xi)$.

The construction of the correct boundary term for specific boundary
conditions can be a rather complicated task. However, to find a {\it
unique\/} expression for $B$ that is compatible with a number of {\it
different\/} boundary conditions is much more difficult. Ideally, we
would like to have a universal expression for $B$ that holds for all
(physically acceptable) boundary conditions. Nester \cite{11} started a
search for an ideal $B$ in the context of PGT by proposing an
expression for it, compatible with solutions having either flat or
constant curvature asymptotic behavior. Later modifications of this
boundary term were intended to make it valid in more general
gravitational theories and for a larger set of boundary conditions, and
also to improve its covariance properties \cite{12,13,14,15}.

In order to properly define the boundary term, it is necessary to
choose a reference dynamical configuration. This configuration is most
naturally linked to the minimal value of the conserved charge
\cite{20}. Let us denote the difference between any variable $X$ and
its reference value $\bar X$ by $\D X=X-\bar X$. In 3D, the boundary
term $B$ is a 1-form. With a suitable set of boundary conditions for
the fields, the proper boundary term reads \cite{13}:
\bsubeq\lab{3.1}
\be
B=(\xi\inn b^i)\D \t_i + \D b^i(\xi\inn\bt_i)
  +(\tilde\nabla_j\xi^i)\D\r_i{^j}
  + \D\om^i{_j}(\xi\inn\br_i{^j})\, ,                      \lab{3.1a}
\ee
where $\tilde\nabla\xi^i:=d\xi^i+(\om^i{_j}+e_j\inn T^i)\xi^j$, and the
relation to our 3D notation is defined by $\om^{ij}=-\ve^{ijk}\om_k$,
$\r_{ij}=-\ve_{ijk}\r^k$. If $\xi$ is asymptotically a Killing vector,
this formula can be simplified by choosing $b^i$ so that $\Lie_\xi b^i$
vanishes on the boundary. In that case, the identity $\Lie_\xi
b^i\equiv\tilde\nabla\xi^i-(\xi\inn\om^i{_j})b^j$ leads to
\be
B=(\xi\inn b^i)\D \t_i + \D b^i(\xi\inn\bt_i)
  +(\xi\inn\om^i{_j})\D\r_i{^j}
  +\D\om^i{_j}(\xi\inn\br_i{^j})\, .                       \lab{3.1b}
\ee
\esubeq
One should note that when triad and connection are independent
dynamical variables, the concept of Killing vector differs from the
usual one, as defined in GR; see, for instance \cite{21}. By our
convention, since $\om^{ij}$ and $\r_{ij}$ are antisymmetric objects,
the summation over $i,j$ in the last two terms goes over $i < j$. After
returning to our 3D notation, we obtain
\be
B=(\xi\inn b^i)\D \t_i + \D b^i(\xi\inn\bt_i)
  +(\xi\inn\om^i)\D\r_i + \D\om^i(\xi\inn\br_i)\, .        \lab{3.2}
\ee

In our calculations of the boundary integrals, we shall use the
Schwarzschild-like coordinates $x^\m=(t,r,\vphi)$. For solutions with
Killing vectors $\pd_t$ and $\pd_\vphi$, the conserved charges are
energy and angular momentum, respectively:
\bsubeq\lab{3.3}
\bea
&&E=\int_{\pd\S}B(\pd_t)
   =\int_{\pd\S} b^i{_0}\D \t_i + \D b^i \bt_{i0}
        +\om^i{_0}\D\r_i + \D\om^i\br_{i0}\, ,             \lab{3.3a}\\
&&M=\int_{\pd\S}B(\pd_\vphi)
   =\int_{\pd\S} b^i{_2}\D \t_i + \D b^i \bt_{i2}
        +\om^i{_2}\D\r_i + \D\om^i\br_{i2}\, ,             \lab{3.3b}
\eea
\esubeq
where $\pd\S$ is a circle (which may be located at infinity), described
by coordinate $\vphi$.

\section{Conserved charges in 3D gravity}
\setcounter{equation}{0}

In 4D, the expression \eq{3.1a} for $B$ was shown to be valid for
general gauge theories of gravity, such as PGT or metric-affine gravity (MAG), and for a large
set of known solutions with different boundary conditions
\cite{11,12,13,14,15}. Here, we wish to verify the correctness of $B$
in 3D gravity by evaluating the conserved charges for several solutions
in either 3D gravity with torsion or topologically massive gravity.

\subsection{BTZ black hole in 3D gravity with or without torsion}

The BTZ black hole \cite{22}, a well-known solution of Einstein's 3D
gravity in the AdS sector (with $\L = -1/\ell^2$), is also a solution
of 3D gravity with torsion \cite{22,16}. In the Schwarzschild-like
coordinates, the black hole metric is given by
\bea
&&ds^2=N^2dt^2-N^{-2}dr^2-r^2(d\vphi+N_\vphi dt)^2\, ,     \nn\\
&&N^2=\left(-8Gm+\frac{r^2}{\ell^2}+\frac{16G^2J^2}{r^2}\right)\, ,
  \qquad N_\vphi=\frac{4GJ}{r^2}\, .                       \nn
\eea

Our approach to 3D gravity with torsion is based on the PGT formalism;
see \cite{16}. Given the metric, we choose the triad field to have the
simple, ``diagonal" form:
\bsubeq\lab{4.1}
\be
b^0=Ndt\, ,\qquad b^1=N^{-1}dr\, ,\qquad
b^2=r\left(d\vphi+N_\vphi dt\right)\, ,                    \lab{4.1a}
\ee
while the solution for Cartan's connection reads
\be
\om^i=\tom^i+\frac{p}{2}\,b^i\, ,                          \lab{4.1b}
\ee
where $\tom^i$ is Riemannian connection,
\be
\tom^0=-Nd\vphi\, ,\qquad \tom^1=N^{-1}N_\vphi dr\, ,\qquad
\tom^2=-\frac{r}{\ell^2}dt-rN_\vphi d\vphi\, ,
\ee
\esubeq
and $p=(\a_3\L+\a_4 a)/(\a_3\a_4-a^2)$. Equations \eq{4.1} define the
analogue of the BTZ black hole in Riemann-Cartan spacetime.

The black hole solution \eq{4.1} possesses two Killing vectors,
$\xi_{(0)}=\pd_t$ and $\xi_{(2)}=\pd_\vphi$, which leave the triad
field $b^i$ form invariant, $\Lie_\xi b^i=0$. As the reference
configuration, we take the solution \eq{4.1} with $m=J=0$ (the black
hole vacuum). The conserved charges are defined in \eq{3.3}, with the
field momenta $\t_i$ and $\r_i$ given in \eq{2.2c}. Using the
asymptotic behavior of $b^i$ and $\om^i$ (Appendix A), we obtain the
following expressions for the energy and angular momentum,
respectively:
\be
E=m+\frac{\a_3}{a}\left(\frac{mp}{2}
                        -\frac{J}{\ell^2}\right)\, ,\qquad
M=J+\frac{\a_3}{a}\left(\frac{pJ}{2}-m\right)\,.         \lab{4.2}
\ee
The result was derived in \cite{22}$_1$ using Nester's covariant method,
and rederived in \cite{16} using Dirac's canonical formalism.
Recalling the expressions for the black hole entropy and angular
velocity \cite{16}, one can verify the validity of the first law of
black hole thermodynamics.

For $\a_3=\a_4=0$, 3D gravity with torsion reduces to Einstein's 3D
gravity, and we have $E=m$ and $M=J$, as expected.

\subsection{BTZ black hole in \tmgl}

The BTZ black hole is a trivial solution of \tmgl, since the related
Cotton tensor identically vanishes. The solution for $b^i$ and $\om^i$
has the same form as in \eq{4.1} but with $p=0$,
\bsubeq\lab{4.3}
\bea
&&b^0=Ndt\, ,\qquad b^1=N^{-1}dr\, ,\qquad
  b^2=r\left(d\vphi+N_\vphi dt\right)\, ,                  \lab{4.3a}\\
&&\om^i=\tom^i\, ,                                         \lab{4.3b}
\eea
while the Lagrange multiplier is given by \cite{8}
\be
\l^i=\frac{a}{\m\ell^2}b^i\, .                             \lab{4.3c}
\ee
\esubeq

To find the conserved charges, we combine \eq{2.3b} with
\eq{4.3c} to obtain
\be
\t_i=\frac{a}{\m\ell^2}b^i\, ,\qquad
\r_i=2ab_i+\frac{a}{\m}\om^i\, .                           \lab{4.4}
\ee
Note that the solution \eq{4.3} and the covariant field momenta
\eq{4.4} can be obtained from the corresponding expressions  \eq{4.1}
and \eq{2.2c} in 3D gravity with torsion, by implementing the following
limit on parameters:
$$
p\to 0\, ,\qquad \a_3\to \frac{a}{\m}\, ,
          \qquad \a_4\to\frac{a}{\m\ell^2}\, .
$$
Hence, the same limit yields the conserved charges in \tmgl:
\be
E=m-\frac{J}{\m\ell^2}\, ,\qquad  M=J-\frac{m}{\m}\, ,     \lab{4.5}
\ee
in complete agreement with the results found in \cite{2,8,10}. The
first law of black hole thermodynamics is a direct consequence of the
arguments given in the previous subsection.

At the chiral points $\m\ell=\mp 1$, the BTZ charges \eq{4.5} satisfy
the chirality relations $\ell E=\pm M$, independently of the values of
$m$ and $J$. Specifically, for the extreme black holes $J=\mp m\ell$ at
the chiral points $\m\ell=\mp 1$, both energy and angular momentum
vanish:
\be
\m\ell=\mp 1 {~\rm and~} J=\mp m\ell\quad\Ra\quad E=M=0\, .\lab{4.6}
\ee

\subsection{Spacelike stretched black hole in \tmgl}

After introducing a convenient notation, $\L=-1/\ell^2$ and
$\n=\m\ell/3$, the metric of the spacelike stretched black hole can be
written in the form \cite{7,5}
\bsubeq
\be
ds^2=N^2dt^2-B^{-2}dr^2-K^2(d\vphi+N_\vphi dt)^2\, ,       \lab{4.7a}
\ee
where
\bea &&N^2=\frac{(\n^2+3)(r-r_+)(r-r_-)}{4K^2}\, ,\qquad
  B^2=\frac{4N^2K^2}{\ell^2}\, ,                           \nn\\
&&K^2=\frac{r}{4}\left[3(\n^2-1)r+(\n^2+3)(r_++r_-)
                 -4\n\sqrt{r_+r_-(\n^2+3)}\right]\, ,      \nn\\
&&N_\vphi=\frac{2\n r-\sqrt{r_+r_-(\n^2+3)}}{2K^2}\, .     \lab{4.7b}
\eea
\esubeq
The solution is not of constant curvature.

Going over to the PGT formalism \cite{9}, we choose the triad field as
\bsubeq\lab{4.8}
\be
b^0=Ndt\, ,\qquad b^1=B^{-1}dr\, ,\qquad
b^2=K(d\vphi+N_\vphi dt)\, ,                               \lab{4.8a}
\ee
and calculate the corresponding Riemannian connection $\tom^i$:
\bea
&&\tom^0=-\frac{N\n}\ell dt-\frac{2NKK'}{\ell}d\vphi\, ,\qquad
  \tom^1=-\frac{KN_\vphi'}{2N}dr\, ,                        \nn\\
&&\tom^2=-\frac{KN_\vphi\n}\ell dt
        +\frac{K^3N_\vphi'}\ell d\vphi\, .                 \lab{4.8b}
\eea
Finally, the solution for $\l_m$ has the form
\bea
&&\l_0=\frac{2a}{\m\ell^2}\left[
       \left(-\frac 32+2\n^2\right)N dt
       +3(\n^2-1)NK^2N_\vphi d\vphi \right]\, ,            \nn\\
&&\l_1=\frac{2a}{\m\ell^2}\left( -\frac{3}{2}+\n^2\right)B^{-1}dr\,,\nn\\
&&\l_2=\frac{2a}{\m\ell^2}\left[
       \left(\frac 32-2\n^2\right)KN_\vphi dt
  +\left(\frac{3}{2}-2\n^2-3(\n^2-1)N^2\right)Kd\vphi\right]\,.\lab{4.8c}
\eea
\esubeq
The result follows from the general formula
\be
\l_m=2a\m^{-1}\left[ (Ric)_{mn}-\frac{1}{4}\eta_{mn}R\right]b^n\, ,
                                                           \lab{4.9}
\ee
where $(Ric)_{mn}=-\ve^{ij}{_m}R_{ijn}$ is the Ricci tensor, and
$R=-\ve^{ijm}R_{ijm}$ the scalar curvature, see Eqs. (2.5c) and (B.2b)
of Ref. \cite{9}.

Now, using the expressions \eq{2.3b} for $\t_i$ and $\r_i$ and the
asymptotic behavior of $N,B,K$ and $N_\vphi$ (Appendix B), with
reference configuration defined by $r_-=r_+=0$, Eq. \eq{3.3} yields the
energy and angular momentum of the spacelike stretched black hole
\eq{4.8}:
\bea
&&E=\frac{(\n^2+3)}{24G\ell}\left[r_++r_-
    -\frac{1}{\n}\sqrt{r_+r_-(3+\n^2)}\right]\,            \nn\\
&&M=\frac{\n(\n^2+3)}{96G\ell}\left[\left(r_++r_-
  -\frac{1}{\n}\sqrt{r_+r_-(3+\n^2)}\right)^2
  -\frac{5\n^2+3}{4\n^2}(r_+-r_-)^2\right]\, .             \lab{4.10}
\eea
These values coincide with those found in \cite{7,9,5}, and moreover,
they are in agreement with the first law of black hole thermodynamics
\cite{7,5}.  On the other hand, a comparison with the results of Ref.
\cite{10} shows a discrepancy by a factor of $1/2$ in energy.

\subsection{Logarithmic solution of \tmgl}

Let us now consider another genuine solution to \tmgl, the logarithmic
solution at the chiral point $\m\ell=-1$ \cite{23}. The metric of the
solution is stationary and spherically symmetric:
\bsubeq\lab{4.11}
 \be
ds^2=A^2dt^2-B^{-2}r^2-K^2(d\vphi+Cdt)^2\, ,              \lab{4.11a}
\ee where:
 \bea
 &&A=\frac r{K}B\,,\quad
  B=\frac r\ell-\frac{4Gm\ell}{r}\, ,                      \nn\\
&&K^2=r^2+\ell^2L_k\,,\qquad
  C=-\frac{4G\ell m+\ell L_k}{K^2}\, ,                    \nn\\
&&L_k=k\ln\frac{r^2-4Gm\ell^2}{r_0^2}\, .
\lab{4.11b} 
\eea 
\esubeq 
The metric depends on two constants $m$
and $k$, whereas $r_0$ is a normalization factor that does not
influence the values of the conserved charges, but does have an
impact on the geometric properties of the solution. For $k=0$,
\eq{4.11} reduces to the extreme BTZ black hole with $J=-m\ell$,
at the chiral point $\m\ell=-1$.

In the PGT formalism, we start with
\bsubeq\lab{4.12}
\be
b^0=Adt\, ,\qquad b^1=B^{-1}dr\, ,\qquad
b^2=K(d\vphi+C dt)\, ,                                     \lab{4.12a}
\ee
the corresponding Riemannian connection $\om^i=\tom^i$ takes the
form
\bea
&&\tom^0=\frac kKdt-\frac{r^2+\ell^2(k-4Gm)}{\ell K}d\vphi\,,\nn\\
&&\tom^1=-\frac{\ell(4Gm-k+L_k)}{BK^2}dr\, ,               \nn\\
&&\tom^2=\left(\frac kK-\frac K{\ell^2}\right)dt
  +\frac{\ell(4Gm-k+L_k)}Kd\vphi\, ,                       \lab{4.12b}
\eea
and the solution for $\l^m$ reads:
\bea
&&\l^0=-\left[\frac{a}{\ell}A
              +\frac{4a\ell k}{K^2}(A-KC)\right]dt
       +\frac{4a\ell k}{K}d\vphi\, ,                       \nn\\
&&\l^1=-\frac{a}{\ell B}dr\, ,                             \nn\\
&&\l^2=-\left[\frac{a}{\ell}KC
          +\frac{4a\ell k}{K^2}(A-KC)\right]dt
  -\left(\frac{a}{\ell}K-\frac{4a\ell k}{K}\right)d\vphi\,.\lab{4.12c}
\eea
\esubeq
For a detailed derivation of \eq{4.12}, see Appendix C.

Using our basic formula \eq{3.3} for the conserved charges, the
expressions \eq{2.3b} for the field momenta $\t_i,\r_i$, the asymptotic
behavior as described in Appendix D and taking the reference
configuration with $m=k=0$, we find that the energy and angular
momentum of the logarithmic solution \eq{4.12} are:
\be
E=\frac{k}{2G}\, , \qquad   M=-\frac{k\ell}{2G}\, .        \lab{4.13}
\ee

To verify \eq{4.13}, we first used the standard canonical approach and
found that the energy and angular momentum of the logarithmic solution
\eq{4.12} are given by the formulas

\bea
&&E_c=\int_{\pd\S} 2ab^0{_0}\left(\om^0-\frac{1}{2\ell}b^0
  +\frac{1}{2a}\l^0+\frac 1{\ell}b^2-\om^2\right)\, ,      \nn\\
&&M_c=-\int_{\pd\S} 2ab^2{_2}\left(\om^2-\frac{1}{2\ell}b^2
  +\frac 1{2a}\l^2+\frac 1{\ell}b^0-\om^0\right)\, ,       \nn
\eea
which are the same as those describing the AdS sector of \tmgl, taken
at $\m\ell=-1$ \cite{8}. The evaluation of these expressions produces
the same result as in \eq{4.13}. Moreover, since the Hawking
temperature of the logarithmic solution vanishes, $4\pi
T=\left(A^2\right)'|_{r_+}=0$, and the angular velocity is
$\Om=C|_{r_+}=-1/\ell$ (with $r_+^2=4Gm\ell^2$), it follows that the
charges \eq{4.13} satisfy the first law of black hole thermodynamics:
$$
\d E-\Om\d M=\d E+\frac 1\ell \d M=0\, .
$$

The results \eq{4.13} essentially agree with those obtained by Clement
\cite{17}, up to some differences in conventions. Indeed, his angular
momentum and $\Om$ have oposite signs as compared to ours, but they
still satisfy the first law. Our conserved charges also agree with the
results found in \cite{10}, but they differ by an overall factor $3/2$
when compared to \cite{23}.

Let us observe that in parallel with the logarithmic solution at the
chiral point $\m\ell=-1$, there exists another logarithmic solution at
$\m\ell=1$, obtained from \eq{4.12} by changing the sign of $C$. For
$k=0$, the new solution reduces to the extreme BTZ black hole with
$J=m\ell$. An analogous evaluation of its conserved charges leads to
$$
\bar{E}=\frac{k}{2G}\, , \qquad \bar{M}=\frac{k\ell}{2G}\, .
$$
Since now $\bar\Om=1/\ell$, the first law is again satisfied. Note that
this ``antichiral" solution is {\it different\/} from the solution
displayed in Eq. (25) of Ref. \cite{23}.

\section{Concluding remarks}

In this paper, we examined Nester's covariant canonical expression
\eq{3.1} for the conserved charges in 3D gravity. It is shown that the
evaluated energy and angular momentum of the BTZ black hole (in both 3D
gravity with torsion and \tmgl), spacelike stretched black hole and the
logarithmic solution, are in complete agreement with the results
obtained by different methods in \cite{2,7,8,9,16,17}. Moreover, for
each of these solutions, the calculated conserved charges are seen to
satisfy the first law of black hole thermodynamics.

On the other hand, the authors of \cite{10}, working in the context of
\tmgl, obtained the same results as ours in the case of the BTZ black
hole and the logarithmic solution, but their treatment of the spacelike
stretched black hole features certain difficulties. Thus, our approach
based on \eq{3.1} exhibits a wider range of validity.

The Lagrangians considered in the present paper are linear in torsion
and/or curvature.  As a further test of universality of the formula
\eq{3.1} in 3D gravity, it would be interesting to apply it to the
Lagrangians quadratic in torsion and/or curvature. After this paper has
been completed, we started studying the conserved charges in new
massive gravity \cite{24}. As a first result in this direction, we
found that the formula \eq{3.3} gives the correct conserved charges for
the BTZ solution \cite{25}.

\section*{Acknowledgements}

This work was partially supported by the Serbian Science Foundation
under Grant No. 141036.

\appendix

\section{Asymptotics of the BTZ black hole}
\setcounter{equation}{0}

In this appendix, we display the asymptotic form of the BTZ black hole
solution in 3D gravity with torsion and in \tmgl. The subleading terms
are given up to the order that contributes to the values of conserved
charges.

\prg{BTZ black hole in 3D gravity with torsion.} Starting with
\eq{4.1a}, we find the asymptotic form of the triad field:
\bsubeq\lab{A.1}
\bea
&&b^0\sim \left(\frac r\ell-\frac{4Gm\ell}r\right)dt\, ,\qquad
  b^1\sim\left(\frac\ell r+\frac{4Gm\ell^3}{r^3}\right)dr\,,\nn\\
&&b^2\sim \frac{4GJ}r dt+rd\vphi\, .                       \lab{A.1a}
\eea
Recalling that Cartan's connection has the form \eq{4.1b},
\be
\om^i=\tom^i+\frac{p}{2}b^i\, ,                            \lab{A.1b}
\ee
it follows that its asymptotics is determined by \eq{A.1a} and
\bea
&&\tom^0\sim\left(-\frac{r}{\ell}+\frac{4Gm\ell}{r}\right)d\vphi\, ,
  \qquad\tom^1\sim \frac{4GJ\ell}{r^3}dr\, ,               \nn\\
&&\tom^2\sim-\frac{r}{\ell^2}dt-\frac{4GJ}{r}d\vphi\, .    \lab{A.1c}
\eea
\esubeq

\prg{BTZ black hole in \tmgl.} Looking at \eq{4.3}, we see that the
asymptotic form of the basic dynamical variables in \tmgl\ is
determined by the formulas \eq{A.1} taken at $p=0$, where the torsion
vanishes.

\section{Asymptotics of the spacelike stretched black hole}
\setcounter{equation}{0}

Dynamical variables of the spacelike stretched black hole solution
\eq{4.8} are expressed in terms of the functions $N,K$ and $N_\vphi$.
Consequently, the asymptotic behavior of the solution is determined by
the following asymptotic relations:
\bea
N&\sim& \sqrt{\frac{\n^2+3}{3(\n^2-1)}}
   -\frac{2\n\sqrt{\n^2+3}\left(\n(r_++r_-)
     -\sqrt{(\n^2+3)r_+r_-}\right)}{[3(\n^2-1)]^{3/2}r}\, ,\nn\\[5pt]
K&\sim&\frac{\sqrt{3(\n^2-1)}}{2}r
   +\frac{(\n^2+3)(r_++r_-)-4\n\sqrt{(\n^2+3)r_+r_-}}
         {4\sqrt{3(\n^2-1)}}                               \nn\\[3pt]
    &&-\frac{\left((\n^2+3)(r_++r_-)-4\n\sqrt{(\n^2+3)r_+r_-}\right)^2}
       {16[3(\n^2-1)]^{3/2}r}\, ,                          \nn\\[3pt]
N_\vphi&\sim&\frac{4\n}{3(\n^2-1)r}
   -\frac{2\left(2\n(\n^2+3)(r_++r_-)-(5\n^2+3)\sqrt{(\n^2+3)r_+r_-}\right)}
     {[3(\n^2-1)]^2r^2}\, .                                \nn
\eea

\section{Derivation of the logarithmic solution}
\setcounter{equation}{0}

To derive the logarithmic solution \eq{4.12}, we start with the triad
field \eq{4.12a}, and calculate the Riemannian connection
$\om^i=\tom^i$:
\be
\tom^0=-\b b^0-\g b^2\, ,\qquad \tom^1=-\b b^1\, ,\qquad
\tom^2=-\a b^0+\b b^2\, ,                                  \nn
\ee
where
\bea
&&\a:=\frac{BA'}{A}
     =\frac{r^2+\ell^2(4Gm-k+2L_k)}{\ell K^2}\, ,          \nn\\[3pt]
&&\b:=\frac{BKC'}{2A}=\frac{\ell(4Gm-k+L_k)}{K^2}\, ,      \nn\\[3pt]
&&\g:=\frac{BK'}K=\frac{r^2+\ell^2(k-4Gm)}{\ell K^2}\, .   \nn
\eea
This result implies \eq{4.12b}. Note that the coefficients $\a,\b$ and
$\g$ are not independent:
$$
\a-\b=\frac{1}{\ell}\, ,\qquad \b+\g=\frac{1}{\ell}\, .
$$

In the next step, we calculate the Riemann curvature:
\bea
&&\tR_0=-\left(\frac 1{\ell^2}-\frac{2k}{K^2}\right)b^1b^2
        +\frac{2k}{K^2}b^0b^1\,,                           \nn\\
&&\tR_1= -\frac{1}{\ell^2}b^2b^0\, ,                       \nn\\
&&\tR_2=-\left(\frac1{\ell^2}
        +\frac{2k}{K^2}\right)b^0b^1-\frac{2k}{K^2}b^1b^2\,.\nn
\eea
The solution for $\l^m$, based on Eq. \eq{4.9}, takes the form
\bea
&&\l^0=-2a\ell\left[\frac{1}{2\ell^2}b^0
       +\frac{2k}{K^2}\left(b^0-b^2\right)\right]\, ,      \nn\\
&&\l^1=-2a\ell\left(\frac{1}{2\ell^2}b^1\right)\,,         \nn\\
&&\l^2=-2a\ell\left[\frac{1}{2\ell^2}b^2
       +\frac{2k}{K^2}\left(b^0-b^2\right)\right]\, ,      \nn
\eea
which implies \eq{4.12c}.

In order to show that \eq{4.12} is indeed a solution of \tmgl, we
calculated the Cotton 2-form $C_i=(\m/2a)\nab\l_i$,
\bea
&&C_0=\frac{2k}{\ell K^2}\left(b^0b^1+b^1b^2\right)\, ,    \nn\\[3pt]
&&C_1=0\, ,\qquad C_2=-C_0\, ,                             \nn
\eea
and verified the basic field equation of \tmgl:
$$
2\tR_i+\frac{1}{\ell^2}\ve_{ijk}b^jb^k+2\m^{-1}C_i=0\, .
$$

\section{Asymptotics of the logarithmic solution}
\setcounter{equation}{0}

The logarithmic solution \eq{4.12} at $\m\ell=-1$ is determined by
the functions $A,B,K$ and $C$. Using  their asymptotic behavior:
\bea &&L_k\sim 2k\ln\frac{r}{r_0}\, ,\qquad
  B\sim\frac{r}{\ell}-\frac{4Gm\ell}{r}\, ,                \nn\\[3pt]
&&K\sim r+\frac{\ell^2L_k}{2r}\, ,\qquad
  C\sim-\frac{4Gm\ell+\ell L_k}{r^2}\, ,                   \nn\\[3pt]
&&A\sim \frac{r}{\ell}-\frac{4Gm\ell+\ell L_k/2}{r}\, .    \nn
\eea we can find the asymptotic form of all the fields needed to
calculate the conserved charges: \bsubeq \bea &&b^0\sim
  \left(\frac{r}{\ell}-\frac{4Gm\ell+\ell L_k/2}{r}\right)dt\, ,\nn\\
&&b^2\sim
  \left( r+\frac{\ell^2L_k}{2r}\right)d\vphi
   -\frac{4Gm\ell+\ell L_k}{r}dt  \, ,
\eea
\bea
&&\tom^0=\frac{k}{r}dt
        -\frac{r^2+\ell^2(k-4Gm-L_k/2)}{\ell r}d\vphi\, ,  \nn\\
&&\tom^2=\left(-\frac{r}{\ell^2}+\frac{2k-L_k}{2r}\right)dt
  +\frac{\ell(4Gm-k+L_k)}{r}d\vphi\, ,
\eea \bea &&\l^0=-a\left(\frac{r}{\ell^2}-\frac{4Gm+L_k/2}{r}
               +\frac{4k}{r}\right)dt
               +\frac{4a\ell k}{r}d\vphi\, ,               \nn\\
&&\l^2=a\left(\frac{4Gm+L_k}{r}+\frac{4k}{r}\right)dt
       -a\ell\left(\frac{r}{\ell^2}+\frac{L_k}{2r}
                    -\frac{4k}{r}\right)d\vphi\, .
\eea
\esubeq

Similarly, one can find the asymptotic form of the logarithmic solution
at $\m\ell=1$.

\end{document}